\documentclass[smallextended]{svjour3}       
\smartqed  
\usepackage{graphicx,epsfig}

\begin{document}

\title{
Quantum Otto heat engines on XYZ spin working medium with DM and KSEA interactions:
Operating modes and efficiency at maximal work output
}

\author{Elena~I.~Kuznetsova\and~M.~A.~Yurischev\and~Saeed~Haddadi
}


\institute{
E.~I.~Kuznetsova \at
Federal Research Center of Problems of Chemical Physics and Medicinal Chemistry,
Russian Academy of Sciences,
Chernogolovka 142432, Moscow Region, Russia\\     
              \email{kuznets@icp.ac.ru}           
           \and
M.~A.~Yurischev
\at
Federal Research Center of Problems of Chemical Physics and Medicinal Chemistry,
Russian Academy of Sciences,
Chernogolovka 142432, Moscow Region, Russia\\     
\email{yur@itp.ac.ru}
           \and
S. Haddadi \at 
School of Physics, Institute for Research in Fundamental Sciences (IPM), P.O. Box
19395-5531, Tehran, Iran\\
and\\
Saeed's Quantum Information Group, P.O. Box 19395-0560, Tehran, Iran\\
              \email{saeed@ssqig.com}           
}

\date{Received:}

\titlerunning{
Quantum Otto heat engines on XYZ spin working medium
 with DM
 and
 KSEA
}
\maketitle

\begin{abstract}
The magnetic Otto thermal machine based on a two-spin-1/2 XYZ working fluid in the
presence of an inhomogeneous magnetic field and antisymmetric Dzyaloshinsky--Moriya
(DM) and symmetric Kaplan--Shekhtman--Entin-Wohlman--Aharony (KSEA) interactions is
considered.
Its possible modes of operation are found and classified.
The efficiencies of engines at maximum power are estimated for various choices of
model parameters.
There are cases when these efficiencies exceed the Novikov value.
New additional points of local minima of the total work are revealed and the mechanism
of their occurrence is analyzed.
\end{abstract}


\section{Introduction}
\label{sect:Intro}
In 1955 Prokhorov and Basov \cite{BP55,BP55a,BP55c}, and later Bloembergen \cite{B56}
proposed a three-level maser scheme with electromagnetic pump to obtain population
inversion, which could lead to negative absorption.
This scheme turned out to be very effective and was successfully implemented in masers
\cite{SFS57,ZKMP58,ZKMP58a} (and then in laser \cite{M60}).
Shortly after, Scovil and Schulz-DuBois came to a conclusion that ``three-level masers
can be regarded as heat engines'', and showed that ``the limiting efficiency of a
3-level maser is that of a Carnot engine'' \cite{SSDB59} (see also
\cite{GSS67,GK94,GK96,LKAS17,GGKNLMSK18,S20}).
The induced (stimulated) emission in such a picture plays a role of the work output of
heat engine, which operates between a hot pump temperature and a low relaxation bath
temperature.
So a three-level maser, as interpreted by Scovil and Schulz-DuBois, was the first
example of a quantum heat engine and has become an important step in the development of
quantum thermodynamics.

Quantum thermodynamics, which grew out of the classical Carnot theory \cite{C1824}, is
based on the quantum-mechanical principles and deals with
conditions of conservation and conversion of such forms of energy as heat and
mechanical work \cite{VA16,BCGAA18,DC19,GMNK19,MAD22} (for a historical review see,
e.g., \cite{AK18}).
Its important branch is the study of quantum cyclic heat engines that produce work
using quantum matter as a working medium.
There are many thermodynamic cycles \cite{QLSN07,Q09}.
One of the most known among them is the Carnot cycle. 
It consists of isentropic compression and expansion and isothermal heat addition and
rejection.
All the processes that compose the ideal Carnot engine can be reversed, in which case
it becomes a heat pump or refrigerator.
The Carnot cycle provides an upper limit to the efficiency that any thermodynamic
engine can achieve when converting heat to work, or vice versa.

Another important cycle is the Otto one.
It is an idealized thermodynamic cycle that describes the operation of a spark-ignited
piston engine in automobiles.
The Otto cycle consists of four processes (strokes): two adiabatic ones, where there
is no heat exchange and two isochoric ones, where there is no work exchange.
Below, we will study {\em magnetic} quantum Otto cycles in which the ``expansion'' and
``compression'' of energy levels of the thermally isolated working fluid are performed
during isochoric processes where work is the change in the average energy due to a
change in external control parameters of the Hamiltonian of the system.

The magnetic Otto cycles and heat machines operating on spin quantum fluid have been
studied by many researches (see Ref.~\cite{PNCV20} and references therein).
As a working substance one takes spin magnetic systems with Heisenberg pair and
multi-spin non-local collective interactions. 
Much attention has been paid to cases when spin working medium involves
Dzyaloshinsky-Moriya (DM) couplings \cite{Z08,ZZ17,AM21}.
However, we are motivated to extend such studies and include in the consideration also
Kaplan--Shekhtman--Entin-Wohlman--Aharony (KSEA) interactions \cite{Y20,FY21}, which
are symmetric in contrast to the DM ones.

The structure of the paper is as follows.
In Sect.~\ref{prelim}, we briefly review quantum Otto cycles composed of two
quantum adiabatic stages and two isochoric coupling to thermal baths (reservoirs).
In Sect.~\ref{sect:model}, we describe the model of working medium used. 
Sect.~\ref{sect:res} is devoted to the description of results obtained and their
discussion.
Finally, our main results are summarized in Sect~\ref{sect:concl}.

\section{Preliminaries}
\label{prelim}
Before we start presenting our results, we should provide some necessary definitions
and expressions used in this paper.

Let there be a system with Hamiltonian $H$, and its density operator $\rho$ satisfies,
say, the quantum Liouville--von Neumann or Lindblad master equation, or has a thermal
equilibrium Gibbs form.
Here we will consider the latter case, that is
\begin{equation}
   \label{eq:rho}
   \rho=\frac{1}{Z}\exp(-\beta H),
\end{equation}
where $Z={\rm Tr}\exp(-\beta H)$ is the partition function and $\beta=1/k_BT$, wherein
$T$ is the temperature, and the Boltzmann constant $k_B$ is assumed to be equal to one
for simplicity.
The operator $\rho$ satisfies the following conditions:
$\rho^\dagger=\rho$, $\rho\ge0$, and ${\rm Tr}\rho=1$.
Next, $F=-T\ln Z$ is the Helmholtz free energy and $S=-\partial F/\partial T$ denotes
the entropy of the system.

The internal energy of the system is given as (see, e.g., \cite{DC19})
\begin{equation}
   \label{eq:U}
   U=\langle H\rangle=\sum_np_nE_n,
\end{equation}
where $E_n$ are the energy levels and the density-matrix eigenvalues
\begin{equation}
   \label{eq:p_n}
   p_n(T)=\frac{1}{Z(T)}\exp(-E_n/T)
\end{equation}
represent the occupation probabilities of energy levels at the temperature $T$.
From here, in accord with the first law of thermodynamics, it follows that during
infinitesimal process the energy change equals \cite{K04}
\begin{equation}
   \label{eq:dU}
   dU=\sum_n(E_ndp_n+p_ndE_n)=\delta Q+\delta W,
\end{equation}
where
\begin{equation}
   \label{eq:dQ}
   \delta Q=\sum_nE_ndp_n
\end{equation}
equals the heat transferred and
\begin{equation}
   \label{eq:dW}
   \delta W=\sum_np_ndE_n
\end{equation}
is the work done.

Note that positive heat, $Q>0$, means that heat is transferred to the working body,
and its negative sign $Q<0$ means that heat, on the contrary, leaves the body.
Similarly for the work.
Positive amount of work, $\delta W>0$, corresponds to the work done on a given body by
external forces, while negative work, $\delta W<0$, means that the body itself does
work on some external object.

\begin{figure}[t]
\begin{center}
\epsfig{file=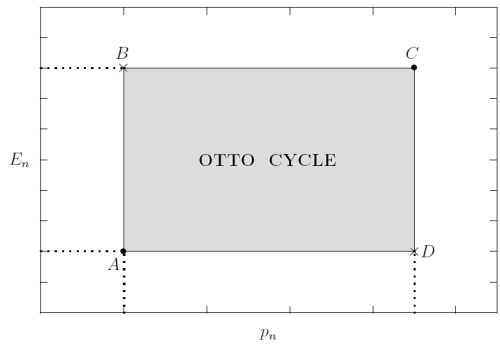,width=6.8cm}
\end{center}
\begin{center}
\caption{
(Color online)
Quantum Otto cycle in the $E_n-p_n$ plane.
The cycle consists of two adiabatic ($AB$ and $CD$) and two isochoric ($BC$ and
$DA$) processes
}
\label{fig:zcycle}
\end{center}
\end{figure}

A cycle of the quantum Otto heat machine consists of four steps (see
Fig.~\ref{fig:zcycle}), namely, two adiabatic processes, where there is no heat
exchange and two so-called isochoric (isomagnetic) ones, where there is no work
exchange \cite{VA16,DC19}.
%
%
All processes are assumed to be sufficiently slow (quasistatic) and the quantum
adiabatic theorem holds
\cite{B27,BF28,M99,QZS05,LG18,IAL21}.
According to this theorem, the level populations are invariant during the course of a
quantum adiabatic process and, consequently, the von Neumann entropy remains
unchanged.
On the other hand, the classical adiabatic process is in equilibrium and
characterized by a certain temperature at any given time, but not necessarily
require that the occupation probabilities remain constant.
The corresponding temperatures can be found, for example, from the Gibbs entropy
invariance condition \cite{PNCV20,SB21}.
Notice that the classical adiabatic theorem is a sequence of quantum adiabatic
theorem, but the converse is not true in general.

Next, the cycle node with the lowest temperature $T_c$ can naturally be referred to
the cold bath, and the node with the highest temperature $T_h$ to the hot one.
Accordingly, the heat from or to the cold and hot baths will be denoted as $Q_c$ and
$Q_h$, respectively.

Let us now consider in detail the Otto cycle shown in Fig.~\ref{fig:zcycle}.
It includes four following strokes.

{\em First stroke}~($AB$).
The working medium at thermal equilibrium with the cold bath in $A$ at the temperature
$T_A=T_c$ is isolated
from thermal reservoir and undergoes an adiabatic compression (magnetization).
Energy level parameters (spacings between energy levels) are
increased, but the occupation probabilities stay unchanged.
The work $W_{A\to B}=W_{in}$ is performed on the working medium during this step:
\begin{equation}
   \label{eq:W_AB}
   W_{A\to B}=\sum_n\int_A^Bp_ndE_n=\sum_np_n^A(E_n^f-E_n^i),
\end{equation}
where $E_n^i$ and $E_n^f$ are the initial and final values of energy levels,
respectively, and $p_n^A=p_n(T_A)$ is the occupation probability by the temperature at
the point $A$.

{\em Second stroke}~($BC$).
The system is brought into thermal contact with the hot bath $C$ under unchanged its
energy structure.
This process is irreversible, and the occupation probabilities change to new
equilibrium values.
Only heat $Q_{B\to C}=Q_h$ is transformed in this step:
\begin{equation}
   \label{eq:Q_BC}
   Q_{B\to C}=\sum_n\int_B^CE_n^fdp_n=\sum_nE_n^f(p_n^C-p_n^B),
\end{equation}
where $p_n^B=p_n(T_A)$ and $p_n^C=p_n(T_C)$ are initial and final values of the
occupation probabilities.

{\em Third stroke}~($CD$).
This is another adiabatic (demagnetization) process reducing the energy gaps to
initial values.
Here, the external control parameters of system are changes back to the initial values
and the occupation probabilities remain fixed.
Only work $W_{C\to D}=W_{out}$ is performed by working medium and no heat is
exchanged:
\begin{equation}
   \label{eq:W_CD}
   W_{C\to D}=\sum_n\int_C^Dp_ndE_n=\sum_np_n^C(E_n^i-E_n^f).
\end{equation}

{\em Fourth stroke}~($DA$).
The system is brought into thermal contact with the cold bath at node $A$.
Again, no work is done, only heat $Q_c$ is rejected during this isochoric process:
\begin{equation}
   \label{eq:Q_DA}
   Q_{D\to A}=\sum_n\int_D^AE_n^idp_n=\sum_nE_n^i(p_n^A-p_n^D).
\end{equation}

\begin{figure}[t]
\begin{center}
\epsfig{file=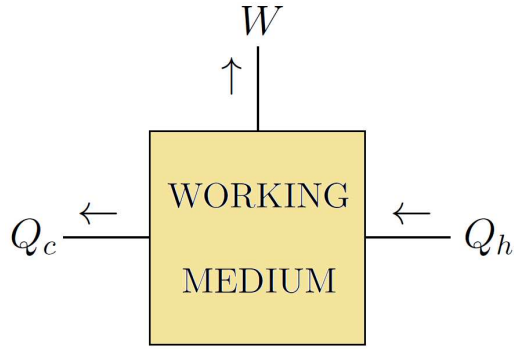,width=6.8cm}
\end{center}
\begin{center}
\caption{
(Color online)
Schematic layout of a heat engine.
The arrows show the energy flows
}
\label{fig:zsch}
\end{center}
\end{figure}

Since energy is conserved in a cyclic process, the balance condition is satisfied:
\begin{equation}
   \label{eq:WQ}
   W_{A\to B}+Q_{B\to C}+W_{C\to D}+Q_{D\to A}=0
\end{equation}
or
\begin{equation}
   \label{eq:WQ1}
   W+Q_h+Q_c=0,
\end{equation}
where 
\begin{equation}
   \label{eq:W}
   W=W_{A\to B}+W_{C\to D}
\end{equation}
is the total work.
If $W<0$, the thermodynamical machine produces mechanical work $|W|=Q_h+Q_c$
with energy absorption $Q_h>0$ and energy release $Q_c<0$,
i.e., corresponds to a heat engine; see Fig.~\ref{fig:zsch}.
%
%
The resulting work $W$ is performed during the exchange of heat $Q_h$ and $Q_c$
between the working fluid and the hot and cold baths.

Instead of drawing, we will further depict the engine as
$\{\circ\!\leftarrow\uparrow\leftarrow\!\bullet\}$
(where the filled circle represents a hot bath and the open circle represents cold
bath) or just like $\{\leftarrow\uparrow\leftarrow\}$
(left - cold bath, right- hot bath).

The efficiency of heat engine is defined as
\begin{equation}
   \label{eq:eta}
   \eta=\frac{|W|}{Q_h}.
\end{equation}
In particular, the efficiency of an ideal Carnot cycle is given by the well-known
expression
\begin{equation}
   \label{eq:eta_Carnot}
   \eta_C=1-\frac{T_c}{T_h},
\end{equation}
which is the upper bound for any thermodynamic cycles.

A heat engine acts by transferring energy from a warm region to a cool region of space
and, in the process, converting some of that energy to mechanical work.
The cycle may also be reversed.
Then the system may be worked upon by an external force,
and in the process, it can transfer thermal energy from a cooler system to a warmer
one, thereby acting as a refrigerator or heat pump
($\{\rightarrow\downarrow\rightarrow\}$) rather than a heat engine.
In this case, the operation of a heat machine is characterized by a coefficient of
performance (CoP), which is defined as
\begin{equation}
   \label{eq:CoP}
   {\rm CoP}=\frac{Q_c}{W}.
\end{equation}

This completes the preliminary section, and now we move on to the description of
the working substance.

\section{Working medium}
\label{sect:model}
As a working medium, we consider a two-site spin-1/2 system with the Hamiltonian
\begin{eqnarray}
   \label{eq:H}
   H&=&
	 J_x\sigma_1^x\sigma_2^x + J_y\sigma_1^y\sigma_2^y + J_z\sigma_1^z\sigma_2^z
	 +D_z(\sigma_1^x\sigma_2^y-\sigma_1^y\sigma_2^x)
	 +{\rm \Gamma}_z(\sigma_1^x\sigma_2^y+\sigma_1^y\sigma_2^x)
	 \nonumber\\
   &+&B_1\sigma_1^z + B_2\sigma_2^z,
\end{eqnarray}
where $\sigma_i^\alpha$ ($i=1,2$; $\alpha=x,y,z$) are the Pauli matrices,
$B_1$ and $B_2$ the $z$-components of external magnetic fields applied at the first
and second qubits respectively, ($J_x$,$J_y$,$J_z$) the vector of interaction
constants of the Heisenberg part of interaction, $D_z$ the $z$-component of
Dzyaloshinsky vector, and ${\rm\Gamma}_z$ the strength of KSEA interaction.
Thus, this model contains seven real independent parameters: $B_1$, $B_2$, $J_x$,
$J_y$, $J_z$, $D_z$, and ${\rm\Gamma}_z$.

In open matrix form, the Hamiltonian (\ref{eq:H}) reads
\begin{equation}
   \label{eq:Hm}
   H=
	 \left(
      \begin{array}{cccc}
      J_z+B_1+B_2&.&.&J_x-J_y-2i{\rm \Gamma}_z\\
      .&-J_z+B_1-B_2\ &J_x+J_y+2iD_z&.\\
      .&J_x+J_y-2iD_z\ &-J_z-B_1+B_2&.\\
      J_x-J_y+2i{\rm \Gamma}_z&.&.&J_z-B_1-B_2
      \end{array}
   \right)
\end{equation}
with the dots which are put instead of zero entries.
This Hermitian matrix has X form.
Its eigenvalues are equal to
\begin{equation}
   \label{eq:Ei}
   E_{1,2}=J_z\pm R_1,\qquad E_{3,4}=-J_z\pm R_2,
\end{equation}
where
\begin{equation}
   \label{eq:R1R2}
   R_1=[(B_1+B_2)^2+(J_x-J_y)^2+4{\rm\Gamma}_z^2]^{1/2},\quad
   R_2=[(B_1-B_2)^2+(J_x+J_y)^2+4D_z^2]^{1/2}.
\end{equation}
Thus, the energy spectrum of the working medium consists of two pairs of levels with
energy shifts $R_1$ and $R_2$.
Therefore, instead of seven parameters, the spectrum is determined by only three
quantities: $J_z$, $R_1$, and $R_2$.
Note that ${\rm\Gamma}_z$ occurs only in $R_1$, while $D_z$ only in
$R_2$, i.e., $R_1$ is the effective ${\rm\Gamma}_z$-parameter, and $R_2$, on the
contrary, is the parameter determined by $D_z$.

The Gibbs density matrix is given by Eq.~(\ref{eq:rho}) and the partition function
$Z=\sum_n\exp(-\beta E_n)$ for the considered model is expressed as
\begin{equation}
   \label{eq:Z}
   Z=2[e^{-\beta J_z}\cosh(\beta R_1)+e^{\beta J_z}\cosh(\beta R_2)].
\end{equation}
Therefore, the Gibbs entropy equals
\begin{eqnarray}
   \label{eq:S}
   &&S(T;J_z,R_1,R_2)=
	 -\frac{1}{Z}\biggl[\frac{R_1-J_z}{T}\exp\left(\frac{R_1-J_z}{T}\right)
	 -\frac{R_1+J_z}{T}\exp\left(-\frac{R_1+J_z}{T}\right)
	 \nonumber\\
   &&+\frac{R_2+J_z}{T}\exp\left(\frac{R_2+J_z}{T}\right)
	 -\frac{R_2-J_z}{T}\exp\left(-\frac{R_2-J_z}{T}\right)\biggr]+\ln Z.
\end{eqnarray}
On the other hand, using Eq.~(\ref{eq:p_n}), the von Neumann entropy
\begin{equation}
   \label{eq:SvN}
   S=-\langle\ln\rho\rangle=-{\rm Tr}\rho\ln\rho=-\sum_np_n\ln p_n
\end{equation}
again leads to the Gibbs entropy expression (\ref{eq:S}).

Finally, using the general relations (\ref{eq:W_AB})--(\ref{eq:Q_DA}) and also
(\ref{eq:p_n}) and (\ref{eq:Ei}), as well as taking into account the quantum adiabatic
theorem, we arrive at equations for a heat engine with the considered working medium.
For the adiabatic (isentropic) strokes, the equations are given as
\begin{eqnarray}
   \label{eq:W_ABa}
   &&W_{in}=
	 \nonumber\\
   &&\Bigl[(J_z^f-J_z^i)\Bigl(\cosh\frac{R_1^i}{T_c}e^{-J_z^i/T_c}-\cosh\frac{R_2^i}{T_c}e^{J_z^i/T_c}\Bigr)
	 -(R_1^f-R_1^i)\sinh\frac{R_1^i}{T_c}e^{-J_z^i/T_c}
	 \nonumber\\
	 &&-(R_2^f-R_2^i)\sinh\frac{R_2^i}{T_c}e^{J_z^i/T_c}\Bigr]
	 /\Bigl(\cosh\frac{R_1^i}{T_c}e^{-J_z^i/T_c}+\cosh\frac{R_2^i}{T_c}e^{J_z^i/T_c}\Bigr),
\end{eqnarray}
and
\begin{eqnarray}
   \label{eq:W_CDa}
   &&W_{out}=
	 \nonumber\\
   &&\Bigl[(J_z^i-J_z^f)\Bigl(\cosh\frac{R_1^f}{T_h}e^{-J_z^f/T_h}-\cosh\frac{R_2^f}{T_h}e^{J_z^f/T_h}\Bigr)
	 -(R_1^i-R_1^f)\sinh\frac{R_1^f}{T_h}e^{-J_z^f/T_h}
	 \nonumber\\
	 &&-(R_2^i-R_2^f)\sinh\frac{R_2^f}{T_h}e^{J_z^f/T_h}\Bigr]
	 /\Bigl(\cosh\frac{R_1^f}{T_h}e^{-J_z^f/T_h}+\cosh\frac{R_2^f}{T_h}e^{J_z^f/T_h}\Bigr).
\end{eqnarray}
The net work done during a cycle is $W=W_{in}+W_{out}$.

Similarly for the isochoric strokes.
The quantities of heat exchanged between working agent and hot and cold reservoirs,
respectively, are
\begin{eqnarray}
   \label{eq:Q_BCa}
   &&Q_h=
   \Bigl[\Bigl(J_z^f\cosh\frac{R_1^f}{T_h}-R_1^f\sinh\frac{R_1^f}{T_h}\Bigr)e^{-J_z^f/T_h}
   -\Bigl(J_z^f\cosh\frac{R_2^f}{T_h}
	 \nonumber\\
	 &&+R_2^f\sinh\frac{R_2^f}{T_h}\Bigr)e^{J_z^f/T_h}\Bigr]
	 /\Bigl(\cosh\frac{R_1^f}{T_h}e^{-J_z^f/T_h}+\cosh\frac{R_2^f}{T_h}e^{J_z^f/T_h}\Bigr)
	 \nonumber\\
   &&-\Bigl[\Bigl(J_z^f\cosh\frac{R_1^i}{T_c}-R_1^f\sinh\frac{R_1^i}{T_c}\Bigr)e^{-J_z^i/T_c}
   -\Bigl(J_z^f\cosh\frac{R_2^i}{T_c}
	 \nonumber\\
	 &&+R_2^f\sinh\frac{R_2^i}{T_c}\Bigr)e^{J_z^i/T_c}\Bigr]
	 /\Bigl(\cosh\frac{R_1^i}{T_c}e^{-J_z^i/T_c}+\cosh\frac{R_2^i}{T_c}e^{J_z^i/T_c}\Bigr),
\end{eqnarray}
and
\begin{eqnarray}
   \label{eq:Q_DAa}
   &&Q_c=
   \Bigl[\Bigl(J_z^i\cosh\frac{R_1^i}{T_c}-R_1^i\sinh\frac{R_1^i}{T_c}\Bigr)e^{-J_z^i/T_c}
   -\Bigl(J_z^i\cosh\frac{R_2^i}{T_c}
	 \nonumber\\
	 &&+R_2^i\sinh\frac{R_2^i}{T_c}\Bigr)e^{J_z^i/T_c}\Bigr]
	 /\Bigl(\cosh\frac{R_1^i}{T_c}e^{-J_z^i/T_c}+\cosh\frac{R_2^i}{T_c}e^{J_z^i/T_c}\Bigr)
	 \nonumber\\
   &&-\Bigl[\Bigl(J_z^i\cosh\frac{R_1^f}{T_h}-R_1^i\sinh\frac{R_1^f}{T_h}\Bigr)e^{-J_z^f/T_h}
   -\Bigl(J_z^i\cosh\frac{R_2^f}{T_h}
	 \nonumber\\
	 &&+R_2^i\sinh\frac{R_2^f}{T_h}\Bigr)e^{J_z^f/T_h}\Bigr]
	 /\Bigl(\cosh\frac{R_1^f}{T_h}e^{-J_z^f/T_h}+\cosh\frac{R_2^f}{T_h}e^{J_z^f/T_h}\Bigr).
\end{eqnarray}

The presented equations make it possible to investigate the quantum Otto heat engine
in the general and various interesting special cases.
Although, nonclassical correlations are initially present in the quantum state $\rho$
with Hamiltonian (\ref{eq:H}), transition to the diagonal energy representation, where
the heat engine is analyzed, completely destroys any quantumness of correlations.


\section{Results and discussion}
\label{sect:res}
Using the above equations, we will now study the operation of the quantum Otto machine
in different modes.


\subsection{A three-level system}
\label{subsect:R1}
Let us start with a simple case, namely, when $J_z=0$ and one of $R_i$ ($i=1,2$) is
also equal to zero.
Without loss of generality, we set $R_2=0$.
In this case, the energy spectrum consists of three levels:
doublet $E_{1,2}=\pm R_1$ and doubly-degenerate zero-energy level $E_{3,4}=0$.

It is important to note that here the energy levels are invariant under the scale
transformation $E_n^f=qE_n^i$, where $q$ is independent of $n$.
This property is necessary and sufficient condition that the quantum adiabatic theorem
reduces to its classical counterpart \cite{LG18}.
Thus, this is the case when the system is quantum but the adiabaticity condition is
classical, i.e., quantum state at each point of the quantum adiabatic process is the
state of thermal equilibrium with respect to the Hamiltonian at the given point.

On the other hand, the Gibbs entropy (\ref{eq:S}) for the case under discussion
reduces to 
\begin{equation}
   \label{eq:S_R1}
	 S(T/R_1)=2\Bigl[\ln2+\ln\Bigl(\cosh\frac{R_1}{2T}\Bigr)-\frac{R_1}{2T}\tanh\frac{R_1}{2T}\Bigr],
\end{equation}
i.e., it is a function of only one variable.
Then the adiabaticity condition is $R_1/T=const$ and, therefore, adiabatic curves in
the plane $R_1-T$ are straight lines passing through the origin of the coordinate
system.

\begin{figure}[t]
\begin{center}
\epsfig{file=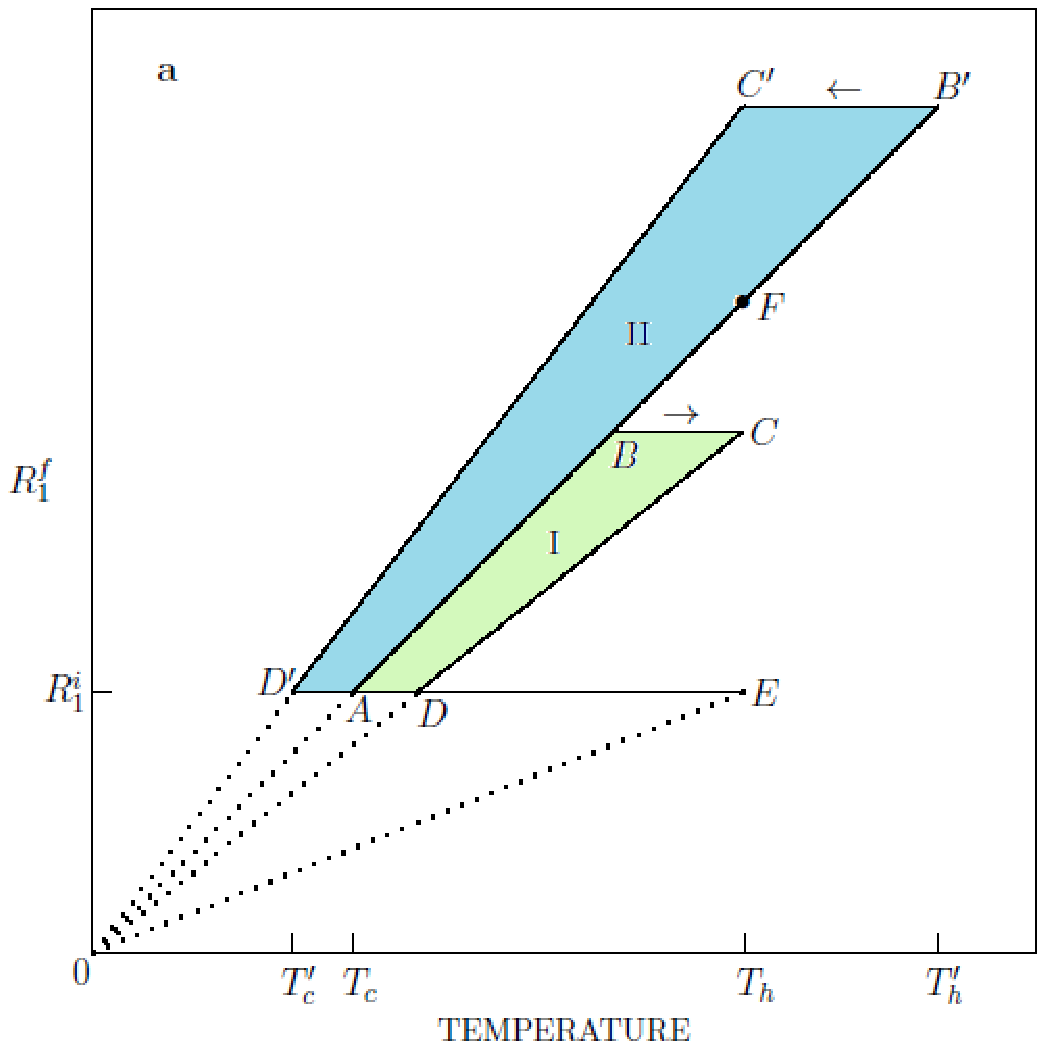,width=6.5cm}
\hspace{-13mm}
\epsfig{file=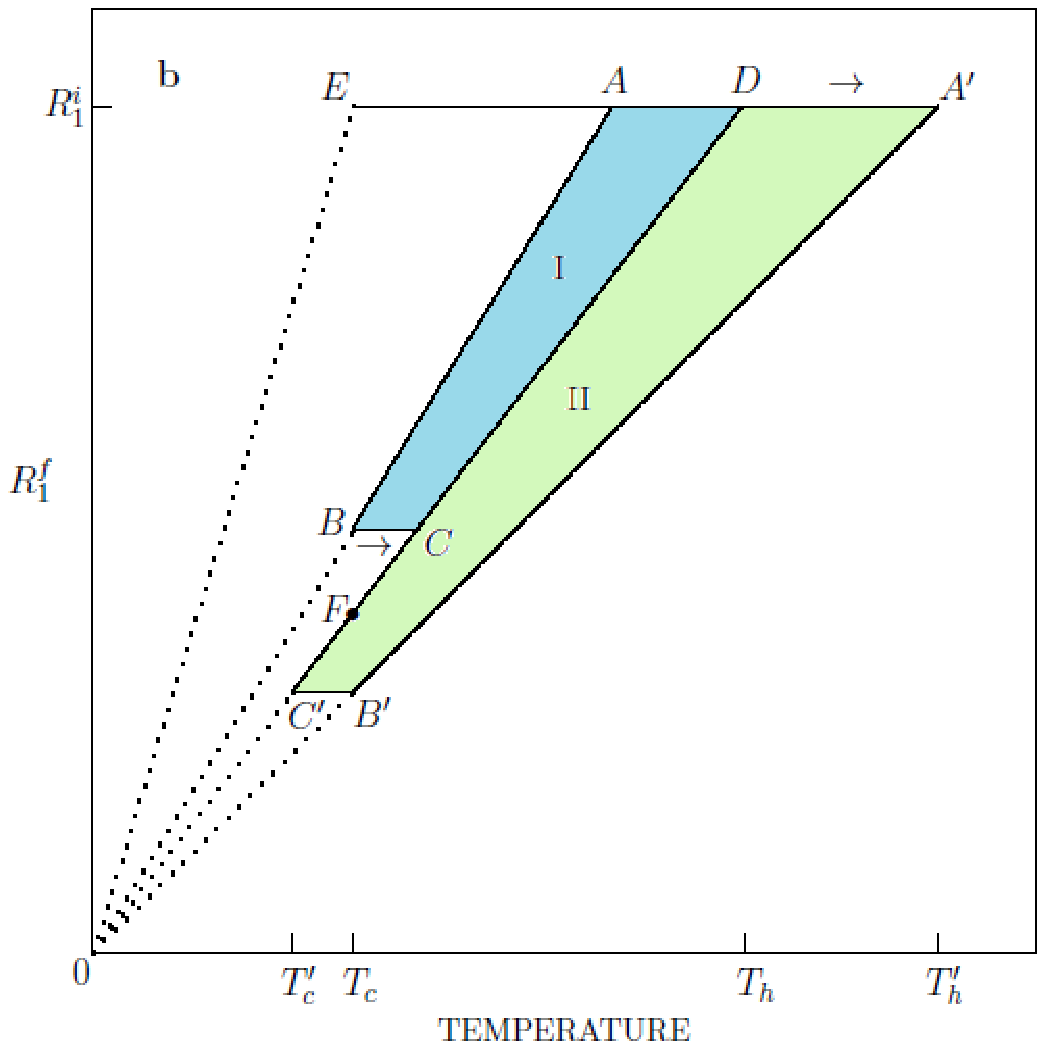,width=6.5cm}
\end{center}
\begin{center}
\caption{
(Color online)
Otto cycles of thermal machine in the case $J_z=R_2=0$ for $R_1^f\ge R_1^i$~(a) and
$R_1^f\le R_1^i$~(b).
The green trapezoids correspond to engine cycles, and the blue ones represent
refrigerator cycles.
Other details are described in the text
}
\label{fig:zcr1ab-c}
\end{center}
\end{figure}

Otto cycles in the plane $R_1-temperature$ are shown in Fig.~\ref{fig:zcr1ab-c}.
%
If the final value of $R_1$ is equal to the initial value, $R_1^f=R_1^i$, then the
cycle contracts into a segment of a horizontal straight line from temperature $T_c$ to
$T_h$ ($AE$ in Fig.~\ref{fig:zcr1ab-c}a and $ED$ in Fig.~\ref{fig:zcr1ab-c}b).

%
\begin{figure}[t]
\begin{center}
\epsfig{file=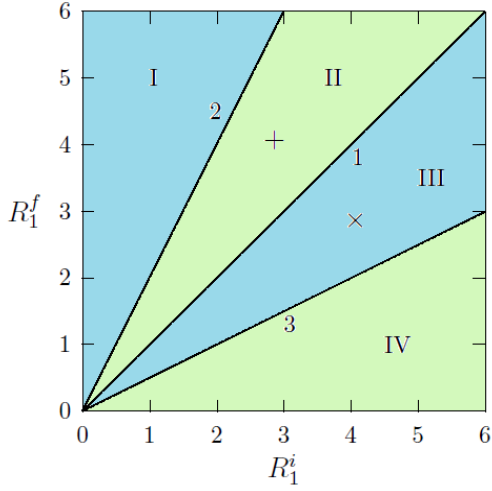,width=8.86cm}
\end{center}
\begin{center}
\caption{
(Color online)
Regions of different operating modes (regimes) in the plane ($R_1^i,R_1^f$) for
the quantum Otto thermal machine with nonzero $R_1$ and $J_z=R_2=0$ by $T_c=1$ and
$T_h=2$.
Here, 1 is the diagonal straight line $R_1^f=R_1^i$, 2 and 3 are the boundaries
$R_1^f=(T_h/T_c)R_1^i$ and $R_1^f=(T_c/T_h)R_1^i$, respectively.
Regions I and III (blue) correspond to the refrigeration regime, while the regions II
and IV (green) represent the heat engine.
The `+' symbol has coordinates (2.86075,4.06548) and marks the position of minimum of
$W$ ($=-0.148615$), and the `$\times$' symbol has mirror coordinates
(4.06548,2.86075) which mark the position of maximum of $W$ ($=0.148615$) in the
region III
}
\label{fig:zr1_2d}
\end{center}
\end{figure}
%

When $R_1^f$ starts to increase, the cycle ABCD appears that goes clockwise and has
adiabatic $AB$ and $CD$ and isochoric $BC$ and $DA$ strokes (green trapezoid I in
Fig.~\ref{fig:zcr1ab-c}a).
The nodes $A$ and $C$ correspond here to the cold and hot reservoirs: $T_A=T_c$ and
$T_C=T_h$.
The adiabaticity conditions allow us to express the temperatures of other two nodes
through the temperatures of the cold and hot reservoirs: $T_B=T_cR_1^f/R_1^i$ and
$T_D=T_hR_1^i/R_1^f$.
Because of this,  the net work performed during the whole cycle is given as
\begin{equation}
   \label{eq:W_R1}
   W=(R_1^f-R_1^i)\Bigl(\tanh\frac{R_1^f}{2T_h}-\tanh\frac{R_1^i}{2T_c}\Bigr).
\end{equation}
This work equals zero, if $R_1^f=R_1^i$ or $R_1^f=(T_h/T_c)R_1^i$.

\begin{figure}[t]
\begin{center}
\epsfig{file=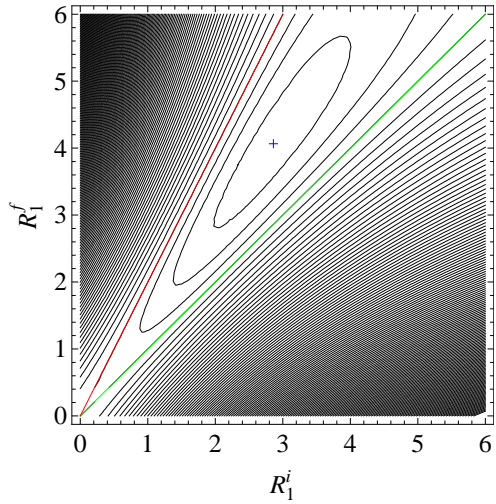,width=6.4cm}
\end{center}
\begin{center}
\caption{
(Color online)
Qualitative structure of isolines for the total work $W(R_1^i,R_1^f)$;
here $J_z=R_2=0$, and bath temperatures $T_c=1$ and $T_h=2$.
The region between the red ($R_1^f=2R_1^i$) and green ($R_1^f=R_1^i$) lines
corresponds to the engine mode.
The symbol ``$+$'' indicates the position of local minimum of the function
$W(R_1^i,R_1^f)$
}
\label{fig:zr1_2dc}
\end{center}
\end{figure}

On the other hand, $Q_h$ and $Q_c$ are given as
\begin{equation}
   \label{eq:QhR1}
   Q_h=R_1^f\Bigl(\tanh\frac{R_1^i}{2T_c}-\tanh\frac{R_1^f}{2T_h}\Bigr).
\end{equation}
and
\begin{equation}
   \label{eq:QcR1}
   Q_c=R_1^i\Bigl(\tanh\frac{R_1^f}{2T_h}-\tanh\frac{R_1^i}{2T_c}\Bigr).
\end{equation}
Both $Q_h$ and $Q_c$ equal zero at the same boundary $R_1^f=(T_h/T_c)R_1^i$ as $W$.
Below this line, $Q_h>0$ and $Q_c<0$.
As a result, the region $0<R_1^f<(T_h/T_c)R_1^i$ corresponds to the heat engine regime;
see green domain II in Fig.~\ref{fig:zr1_2d}.
The structure of isolines of net work $W$ in this domain is depicted in
Fig.~\ref{fig:zr1_2dc}.
%

Taking into account the definition (\ref{eq:eta}) and
Eqs.~(\ref{eq:W_R1})--(\ref{eq:QhR1}), we get the efficiency of the given Otto heat
engine
\begin{equation}
   \label{eq:eta_R1}
   \eta=1 -\frac{R_1^i}{R_1^f}.
\end{equation}
Since $R_1^f<(T_h/T_c)R_1^i$, the value found is less than Carnot's efficiency
(\ref{eq:eta_Carnot}).
This agrees with the Carnot theorem (principle) known from classical thermodynamics.
According to this theorem, no heat engine operating on a cycle between two heat
reservoirs can be more efficient than a reversible heat engine operating between the
same two reservoirs regardless of the working substance employed or the operation
details;
Carnot's efficiency (\ref{eq:eta_Carnot}) is the upper limit that does not depend on
the design of the engine
(see, e.g., \cite{FLS64}, Chapt.~44).

The efficiency (\ref{eq:eta_R1}) is zero at $R_1^f=R_1^i$.
When $R_1^f$ reaches the value of $(T_h/T_c)R_1^i$, the Otto cycle is reduced to a
section of straight line between points $A$ and $F$, as shown in
Fig.~\ref{fig:zcr1ab-c}a.
The efficiency of such a ``cycle'' reaches a Carnot efficiency of $50\%$, however, the
total work $W$, Eq.~(\ref{eq:W_R1}), vanishes here.

Further, if $R_1^f>(T_h/T_c)R_1^i$, the cycle transforms into a trapezoid
$AB^\prime C^\prime D^\prime$ shown in Fig.~\ref{fig:zcr1ab-c}a by a blue region II.
Note first of all, that the direction of such a cycle was changed to opposite.
Moreover, the minimum temperature now is at the node $D^\prime$ and equals
$T_c^\prime=(R_1^i/R_1^f)T_h$, while the maximum one is at node $B^\prime$
and equals $T_h^\prime=(R_1^f/R_1^i)T_c$.
It is clear that $T_c^\prime<T_c$, $T_h^\prime>T_h$ and
$T_h^\prime/T_c^\prime>T_h/T_c$.
Hence, the total work
\begin{equation}
   \label{eq:W_R1-II}
   W=(R_1^f-R_1^i)\Bigl(\tanh\frac{R_1^f}{2T_h}-\tanh\frac{R_1^i}{2T_c}\Bigr)
	 =(R_1^f-R_1^i)\Bigl(\tanh\frac{R_1^i}{2T_c^\prime}-\tanh\frac{R_1^f}{2T_h^\prime}\Bigr)>0.
\end{equation}
The values of heat of cold and hot strokes are given as
\begin{equation}
   \label{eq:QcR1-II}
   Q_c\equiv Q_{D^\prime\to A}=R_1^i\Bigl(\tanh\frac{R_1^f}{2T_h}-\tanh\frac{R_1^i}{2T_c}\Bigr)
	 =R_1^i\Bigl(\tanh\frac{R_1^i}{2T_c^\prime}-\tanh\frac{R_1^f}{2T_h^\prime}\Bigr)>0
\end{equation}
and
\begin{equation}
   \label{eq:QhR1-II}
   Q_h\equiv Q_{B^\prime\to C^\prime}=R_1^f\Bigl(\tanh\frac{R_1^i}{2T_c}-\tanh\frac{R_1^f}{2T_h}\Bigr)
	 =R_1^f\Bigl(\tanh\frac{R_1^f}{2T_h^\prime}-\tanh\frac{R_1^i}{2T_c^\prime}\Bigr)<0.
\end{equation}
This regime corresponds to the refrigerator mode (blue region I in 
Fig.~\ref{fig:zr1_2d}).

We discuss now the cases when $R_1^f$ is less than $R_1^i$.
If $(T_c/T_h)R_1^i<R_1^f<R_1^i$, typical cycle can be represented by a trapezoid
$ABCD$ shown in Fig.~\ref{fig:zcr1ab-c}b as blue region I.
The cycle runs counterclockwise and cold and hot nodes are $B$ and $D$, respectively.
The total work and heat are given by expressions
\begin{equation}
   \label{eq:W_R1-III}
   W=(R_1^i-R_1^f)\Bigl(\tanh\frac{R_1^f}{2T_c}-\tanh\frac{R_1^i}{2T_h}\Bigr)>0,
\end{equation}
\begin{equation}
   \label{eq:QcR1-III}
   Q_c=R_1^f\Bigl(\tanh\frac{R_1^f}{2T_c}-\tanh\frac{R_1^i}{2T_h}\Bigr)>0
\end{equation}
and
\begin{equation}
   \label{eq:QhR1-III}
   Q_h=R_1^i\Bigl(\tanh\frac{R_1^i}{2T_h}-\tanh\frac{R_1^f}{2T_c}\Bigr)<0.
\end{equation}
This is again the cooling mode of Otto's thermal machine:
$\{\rightarrow\downarrow\rightarrow\}$.
For example, the coefficient of performance at the point of maximum total work in this
case (see Fig.~\ref{fig:zr1_2d}, the point marked with the symbol ``$\times$'')
reaches the value ${\rm CoP}=2.37$.

When $R_1^f=(T_c/T_h)R_1^i$, the ``cycle'' is a straight-line section $DF$.
Here, $W=Q_c=Q_h=0$.

Finally, if $R_1^f<(T_c/T_h)R_1^i$ then the cycle is $DA^\prime B^\prime C^\prime$
shown as the green trapezoid II in Fig.~\ref{fig:zcr1ab-c}b.
In this case $W<0$, $Q_c<0$ and $Q_h>0$ and therefore the heat engine regime is
realized here.
In Fig.~\ref{fig:zr1_2d}, the corresponding area is labeled IV and shown in green.

As mentioned above, the efficiency of the discussed Otto engine can reach the upper
limit, namely, the Carnot efficiency.
However, in this case, the total work performed is zero and therefore such an
``engine'' is useless.
It is of interest to find the efficiency of engines at their maximum power (work
per cycle).

In 1957, Novikov \cite{N57} considered a generalized Carnot engine taking into
account the heat loss from the hot bath to the working fluid
($\{\leftarrow\uparrow\leftarrow\triangleleft\}$, where the triangle $\triangleleft$
denotes a lossy heat conductor)
and derived a remarkable formula for the efficiency at maximum power of such an engine
(Eq.~(7) in Ref.~\cite{N57} and Eq.~(6) in Ref.~\cite{N58})
\begin{equation}
   \label{eq:eta_N}
   \eta_{N}=1-\sqrt{T_c/T_h}.
\end{equation}
(In connection with the problem of optimal efficiency of engines, see
Ref.~\cite{VN72}.)
More later, in 18 years, Curzon and Ahlborn \cite{CA75,AC04} (see also
\cite{VLF14,F17}) considered a Carnot engine with losses both from the hot bath to the
working fluid and from the working fluid to the cold bath,
$\{\triangleleft\leftarrow\uparrow\leftarrow\triangleleft\}$,
and obtained the same result for the efficiency.
This gave an impetus to the development of endoreversible thermodynamics
\cite{DC19,H08}.

It turned out that efficiency (\ref{eq:eta_N}) is the benchmark for the efficiency
$\eta_{mp}$ of any real running engine at maximum power.
Therefore, it is interesting to compare the efficiency at maximal power of the Otto
engine with the Novikov efficiency.
The efficiency for the engine operating between bath temperatures $T_c=1$ and
$T_h=2$ at the point with the maximum work done
($|W|=0.148615$, see Figs.~\ref{fig:zr1_2d} and \ref{fig:zr1_2dc}) equals $29.6\%$.
This is less than Carnot's efficiency of $50\%$, but larger than Novikov's efficiency
equaled $29.3\%$.
A similar picture is also valid for other temperatures presented in Table~\ref{tab:1}.
%
\begin{table}[t]
\caption{
Coordinates ($R_1^i$ and $R_1^f$) of a minimum of the work $W$, its value at the
minimum, efficiency at maximum power, and Carnot and Novikov efficiencies by $Tc=1$ and
different values of $T_h$ 
\upshape\upshape}
\label{tab:1}
\begin{tabular}{lccclll}
\hline\noalign{\smallskip}
$T_h$ & $R_1^i$ & $R_1^f$ & $W$ & $\eta_{mp}$ & $\eta_{C}$ & $\eta_{N}$ \\
\noalign{\smallskip}\hline\noalign{\smallskip}
3 & 3.16836  & 5.59152 & -0.454983  & 43.3\% & 66.7\% & 42.3\% \\
2.5 & 3.02699  & 4.83933 & -0.289598  & 37.5\% & 60\% & 36.8\% \\
2   & 2.86075 & 4.06548 & -0.148615 & 29.6\% & 50\% & 29.3\% \\
1.5 & 2.65857 & 3.25929 & -0.044155 & 18.43\% & 33.3\% & 18.35\% \\
\noalign{\smallskip}\hline
\end{tabular}
\end{table}
As seen from Table~\ref{tab:1}, both the useful work and efficiency grow with
increasing the temperature difference of reservoirs.
Note that these $\eta_{mp}$ values are well reproduced by Novikov's formula, and
moreover, they are somewhat greater than it provides.
A similar increase in efficiency at maximum power has recently been obtained for a
photonic engine \cite{SPD20}.

Thus, the Otto thermal machine on a spin working substance with $J_z=0$ and one of the
two $R_1$ and $R_2$ equal to zero can operate either as an engine or as a refrigerator.
The efficiency at maximum output power is limited from above by the Carnot bound,
and from below by the Novikov efficiency.

\subsection{
Two local minima of net work done
}
\label{subsect:R1R2}
In this section, we extend the case described above and consider the parameter $R_2$
as constant, not equal to zero.
So $J_z=0$, $R_2=const$, and $R_1\in[R_1^i,R_1^f]$.

From Eqs.~(\ref{eq:W_ABa}) and (\ref{eq:W_CDa}), it follows that the net work done
during a cycle is given as
\begin{equation}
   \label{eq:W_R1R2a}
   W=(R_1^f-R_1^i)\Bigl[\sinh\frac{R_1^f}{T_h}/\Bigl(\cosh\frac{R_1^f}{T_h}+\cosh\frac{R_2}{T_h}\Bigr)
	 -\sinh\frac{R_1^i}{T_c}/\Bigl(\cosh\frac{R_1^i}{T_c}+\cosh\frac{R_2}{T_c}\Bigr)\Bigr].
\end{equation}
Next, in accordance with Eqs.~(\ref{eq:Q_BCa}) and (\ref{eq:Q_DAa}), the heat $Q_h$
is
\begin{equation}
   \label{eq:QhR1R2a}
   Q_h=
	 \frac{R_1^f\sinh(R_1^i/T_c)+R_2\sinh(R_2/T_c)}{\cosh(R_1^i/T_c)+\cosh(R_2/T_c)}
	 -\frac{R_1^f\sinh(R_1^f/T_h)+R_2\sinh(R_2/T_h)}{\cosh(R_1^f/T_h)+\cosh(R_2/T_h)}
\end{equation}
and
similarly for $Q_c$:
\begin{equation}
   \label{eq:QcR1R2a}
   Q_c=
	 \frac{R_1^i\sinh(R_1^f/T_h)+R_2\sinh(R_2/T_h)}{\cosh(R_1^f/T_h)+\cosh(R_2/T_h)}
	 -\frac{R_1^i\sinh(R_1^i/T_c)+R_2\sinh(R_2/T_c)}{\cosh(R_1^i/T_c)+\cosh(R_2/T_c)}.
\end{equation}

The boundaries separating the regions with $W>0$ and $W<0$ are found from condition
$W=0$.
It is obvious from (\ref{eq:W_R1R2a}), that one boundary is again the diagonal
\begin{equation}
   \label{eq:R1f_w1}
   R_1^f=R_1^i,
\end{equation}
while the other boundary is determined by the relation
\begin{equation}
   \label{eq:R1R2aR1f}
   R_1^f=T_h\ln\Biggl[\frac{1}{1-\gamma}\Biggl(\gamma\cosh\frac{R_2}{T_h}+\sqrt{1+\gamma^2\sinh^2\frac{R_2}{T_h}}\Biggr)\Biggr],
\end{equation}
where
\begin{equation}
   \label{eq:R1R2a_gamma}
   \gamma=\sinh\frac{R_1^i}{T_c}/\Bigl(\cosh\frac{R_1^i}{T_c}+\cosh\frac{R_2}{T_c}\Bigr).
\end{equation}
It is clear that $R_1^f=0$ at $R_1^i=0$.
For small $R_1^i$, the dependence (\ref{eq:R1R2aR1f}) behaves like
\begin{equation}
   \label{eq:R1R2aR1fa}
   R_1^f\approx\kappa R_1^i,
\end{equation}
where
\begin{equation}
   \label{eq:R1R2kappa}
	 \kappa=\frac{T_h}{T_c}\Bigl(\frac{\cosh[R_2/(2T_h)]}{\cosh[R_2/(2T_c)]}\Bigr)^2.
\end{equation}
For $\kappa=1$, these two boundaries touch near small $R_1^i$.
On the other hand, when $R_1^i\to\infty$, the function $R_1^f$ of $R_1^i$,
Eq.~(\ref{eq:R1R2aR1f}), satisfies the linear asymptotic law
\begin{equation}
   \label{eq:R1R2aR1fas}
   R_1^f\approx T_h\ln\Bigl(\frac{\cosh(R_2/T_h)}{\cosh(R_2/T_c)}\Bigr)+\frac{T_h}{T_c}R_1^i.
\end{equation}
Thus, for large $R_1^i$, the values of $R_1^f$ again follow, as in the previous
subsection, a linear dependence $R_1^f=(T_h/T_c)R_1^i$, but now shifted.

\begin{figure}[t]
\begin{center}
\epsfig{file=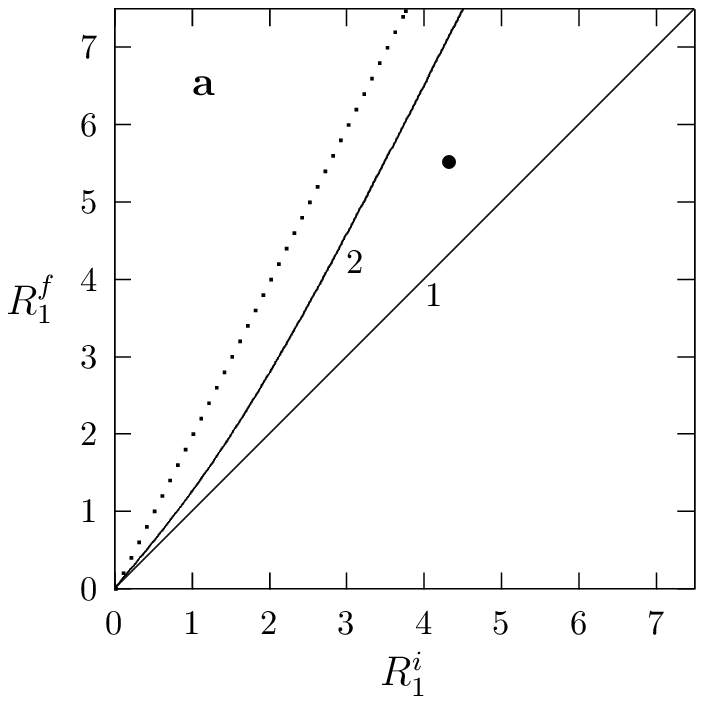,width=5.4cm}
\hspace{4mm}
\epsfig{file=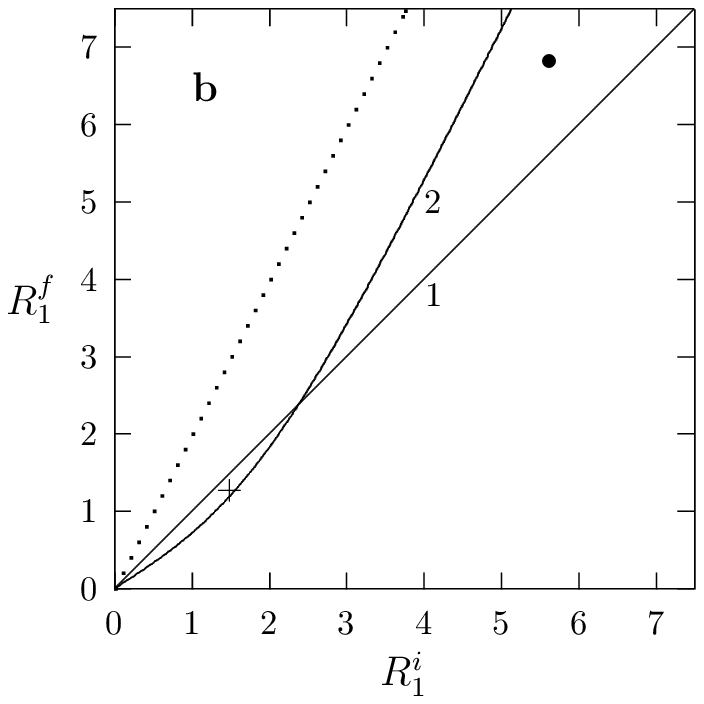,width=5.4cm}
\end{center}
\begin{center}
\caption{
Regions with $W<0$ (between lines 1 and 2) and with $W>0$ (outside the previous
region) in the plane $(R_1^i,R_1^f)$.
Bath temperatures are $T_c=1$ and $T_h=2$.
Dotted line $R_1^f=(T_h/T_c)R_1^i$ is shown for a comparison with the case $R_2=0$.
(a), $R_2=1.8$, the black circle ($\bullet$) has coordinates $(4.32922,5.51837)$ and
indicates the local minimum of work ($W=-0.09977$).
(b), $R_2=2.9$, black circle ($\bullet$) has coordinates $(5.62759,6.82585)$ and
shows the minimum $W=-0.08299$, while the symbol plus (+) marks additional
local minimum ($W=-0.00366$) at the point $(1.31298,1.15942)$
}
\label{fig:zwr1r2a}
\end{center}
\end{figure}

For bath temperatures $T_c=1$ and $T_h=2$, the slope coefficient (\ref{eq:R1R2kappa})
reaches the critical value $\kappa_c=1$ at
$R_2^{(c)}=4\ln\big[\frac{1}{2}\big(\frac{1+\sqrt5}{\sqrt2}+\sqrt{\sqrt5-1}\big)\big]\simeq2.12255$.
When $R_2<R_2^{(c)}$, the engine mode has only one local minimum of the work $W$
(see Fig.~\ref{fig:zwr1r2a}a).
%
Here, both $R_1^i$ and $R_1^f$ are greater than $R_2$, and therefore there is no energy
level crossing.

However, for $R_2>R_2^{(c)}$, the curve 2 forms a loop, inside which the {\em second}
minimum appears (see Fig.~\ref{fig:zwr1r2a}b).
In it, $R_1^i$ and $R_1^f$ are less than $R_2$, which means that again there is no
energy level crossing.
Engine efficiencies, defined by Eq.~(\ref{eq:eta}), in two minima shown in
Fig.~\ref{fig:zwr1r2a}b are $\eta_\bullet=26.3\%$ and $\eta_{+}=0.58\%$.
Both of these values are less than Novikov's efficiency of $29.3\%$.

So, the presence of two minimums of optimal engine operating modes at once is a rather
interesting situation, but it is not yet clear where and how it can be used in
practice.

\subsection{
Three-parameter energy spectrum
}
\label{subsect:JzR1R2}
We now turn to the quantum Otto machine, the working body of which is described by the
Hamiltonian (\ref{eq:H}) with all interactions.
The energy levels are characterized by three parameters $J_z$, $R_1$, and $R_2$.

Let the longitudinal exchange coupling $J_z$ vary within $J_z^i$ and $J_z^f$, while
the parameters $R_1$ and $R_2$ remain unchanged during a cycle, that is
$R_1^i=R_1^f=R_1$ and $R_2^i=R_2^f=R_2$.
Equations~(\ref{eq:W_ABa}) and (\ref{eq:W_CDa}) for the net work done take the form
\begin{eqnarray}
   \label{eq:Wa}
   &&W=(J_z^f-J_z^i)
	 \nonumber\\
   &&\times\Bigl[\Bigl(\cosh\frac{R_1}{T_c}e^{-J_z^i/T_c}-\cosh\frac{R_2}{T_c}e^{J_z^i/T_c}\Bigr)
	 /\Bigl(\cosh\frac{R_1}{T_c}e^{-J_z^i/T_c}+\cosh\frac{R_2}{T_c}e^{J_z^i/T_c}\Bigr)
	 \nonumber\\
   &&-\Bigl(\cosh\frac{R_1}{T_h}e^{-J_z^f/T_h}-\cosh\frac{R_2}{T_h}e^{J_z^f/T_h}\Bigr)
	 /\Bigl(\cosh\frac{R_1}{T_h}e^{-J_z^f/T_h}+\cosh\frac{R_2}{T_h}e^{J_z^f/T_h}\Bigr)\Bigr].
	 \nonumber\\
\end{eqnarray}
The boundaries separating the regions with $W>0$ and $W<0$ are given here like
\begin{equation}
   \label{eq:Jzf_w1}
   J_z^f=J_z^i
\end{equation}
and
\begin{equation}
   \label{eq:Jzf_w}
   J_z^f=\frac{T_h}{T_c}J_z^i+\frac{1}{2}T_h\ln\Biggl(\frac{\cosh(R_1/T_h)\cosh(R_2/T_c)}{\cosh(R_1/T_c)\cosh(R_2/T_h)}\Biggr).
\end{equation}
These straight lines intersect at a point defined by the presented equations.

Taken into account Eq.~(\ref{eq:Q_BCa}), the heat $Q_h$ for the case under
consideration is reduced to
\begin{eqnarray}
   \label{eq:Q_BCab}
   &&Q_h=A_1
   -\Bigl[\Bigl(J_z^f\cosh\frac{R_1}{T_c}-R_1\sinh\frac{R_1}{T_c}\Bigr)e^{-J_z^f/T_c}
   -\Bigl(J_z^f\cosh\frac{R_2}{T_c}
	 \nonumber\\
	 &&+R_2\sinh\frac{R_2}{T_c}\Bigr)e^{J_z^f/T_c}\Bigr]
	 /\Bigl(\cosh\frac{R_1}{T_c}e^{-J_z^f/T_c}+\cosh\frac{R_2}{T_c}e^{J_z^f/T_c}\Bigr),
\end{eqnarray}
where
\begin{eqnarray}
   \label{eq:A1}
   &&A_1=
   \Bigl[\Bigl(J_z^f\cosh\frac{R_1}{T_h}-R_1\sinh\frac{R_1}{T_h}\Bigr)e^{-J_z^f/T_h}
   -\Bigl(J_z^f\cosh\frac{R_2}{T_h}
	 \nonumber\\
	 &&+R_2\sinh\frac{R_2}{T_h}\Bigr)e^{J_z^f/T_h}\Bigr]
	 /\Bigl(\cosh\frac{R_1}{T_h}e^{-J_z^f/T_h}+\cosh\frac{R_2}{T_h}e^{J_z^f/T_h}\Bigr).
\end{eqnarray}
Putting $Q_h=0$, we get the expression for the boundary in an explicit form
\begin{equation}
   \label{eq:Jzi_h}
   J_z^i=\frac{1}{2}T_c\ln\Biggl(\frac{\cosh(R_1/T_c)}{\cosh(R_2/T_c)}\cdot\frac{J_z^f-R_1\tanh(R_1/T_c)-A_1}{J_z^f
	 +R_2\tanh(R_2/T_c)+A_1}\Biggr).
\end{equation}

Another heat, $Q_c$, is equal to
\begin{eqnarray}
   \label{eq:Q_DAab}
   &&Q_c=A_2
   -\Bigl[\Bigl(J_z^i\cosh\frac{R_1}{T_h}-R_1\sinh\frac{R_1}{T_h}\Bigr)e^{-J_z^i/T_h}
   -\Bigl(J_z^i\cosh\frac{R_2}{T_h}
	 \nonumber\\
	 &&+R_2\sinh\frac{R_2}{T_h}\Bigr)e^{J_z^i/T_h}\Bigr]
	 /\Bigl(\cosh\frac{R_1}{T_h}e^{-J_z^i/T_h}+\cosh\frac{R_2}{T_h}e^{J_z^i/T_h}\Bigr),
\end{eqnarray}
where
\begin{eqnarray}
   \label{eq:A2}
   &&A_2=
   \Bigl[\Bigl(J_z^i\cosh\frac{R_1}{T_c}-R_1\sinh\frac{R_1}{T_c}\Bigr)e^{-J_z^i/T_c}
   -\Bigl(J_z^i\cosh\frac{R_2}{T_c}
	 \nonumber\\
	 &&+R_2\sinh\frac{R_2}{T_c}\Bigr)e^{J_z^i/T_c}\Bigr]
	 /\Bigl(\cosh\frac{R_1}{T_c}e^{-J_z^i/T_c}+\cosh\frac{R_2}{T_c}e^{J_z^i/T_c}\Bigr).
\end{eqnarray}
Setting $Q_c=0$, we obtain an explicit expression for the fourth boundary
\begin{equation}
   \label{eq:Jzf_c}
   J_z^f=\frac{1}{2}T_h\ln\Biggl(\frac{\cosh(R_1/T_h)}{\cosh(R_2/T_h)}\cdot\frac{J_z^i-R_1\tanh(R_1/T_h)-A_2}{J_z^i
	 +R_2\tanh(R_2/T_h)+A_2}\Biggr).
\end{equation}

Thus, mathematical tools are ready, and we can proceed to study the operating modes of
a heat engine.

\begin{figure}[t]
\begin{center}
\epsfig{file=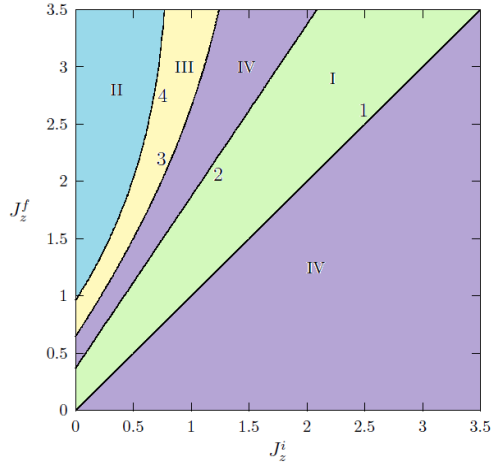,width=6.8cm}
\end{center}
\begin{center}
\caption{
(Color online)
Regions of operation modes in the $J_z^i$-$J_z^f$ plane for quantum Otto thermal
machine with
$R_1=0.7$, $R_2=2$ and bath temperatures
$T_c=1$ and $T_h=1.5$:
I~(green), engine;
II~(blue), refrigerator;
III~(yellow), heater;
IV~(violet), accelerator.
Lines 1--4 are the boundaries separating the listed regions
}
\label{fig:z21}
\end{center}
\end{figure}

Consider, for instance, a spin working medium with parameters $R_1=0.7$ and $R_2=2$,
which is located between the thermal reservoirs at temperatures $T_c=1$ and $T_h=1.5$.
Lines 1, 2, 3 and 4, defined by Eqs.~(\ref{eq:Jzf_w1}), (\ref{eq:Jzf_w}),
(\ref{eq:Jzi_h}) and (\ref{eq:Jzf_c}), divide the plane $J_z^i$-$J_z^f$ into several
regions, as drawn in Fig.~\ref{fig:z21}.
%
%
Regions corresponding to different modes of operation are marked here with Roman
numerals and additionally colored.
The boundaries separating the regions are marked with Arabic numerals 1-4.
It is noteworthy that curves 3 and 4 do not intersect each other and do not intersect
lines 1 and 2.

Finding the signs of $Q_c$, $W$ and $Q_h$ in each such region made it possible to
determine that there are only four different types of regions, see again
Fig.~\ref{fig:z21}.
Firstly, the region I with $Q_c<0$, $W<0$ and $Q_h>0$ naturally corresponds to
engine mode, which is denoted as $\{\leftarrow\uparrow\leftarrow\}$.
Secondly, the region II with $Q_c>0$, $W>0$ and $Q_h<0$, which is identified with a
refrigerator or heat pump, and for clarity we depict it in the form
$\{\rightarrow\downarrow\rightarrow\}$. 
Then the region III, where $W>0$ and both $Q_h$ and $Q_c$ are less than zero; this is
a heater that is represented as  $\{\leftarrow\downarrow\rightarrow\}$.
Finally, the region IV in which $W$ and $Q_h$ are grater than zero while $Q_c<0$,
that is $\{\leftarrow\downarrow\leftarrow\}$;
this is the so-called accelerator or cold-bath heater \cite{GK96,S21,CRAOR22}.

The total work output $W(J_z^i,J_z^f)$, Eq.~(\ref{eq:Wa}), has a local minimum
$W_{min}=-0.030259$ at the point $(0.659225,0.976325)$.
The hot heat (\ref{eq:Q_BCab}) at this point is $Q_h=0.343863$.
Therefore, in accord with Eq.~(\ref{eq:eta}), the efficiency at maximal power equals
$\eta_{mp}=8.8\%$.
This value is less than Novikov's efficiency equal to $18.4\%$.

\begin{figure}[t]
\begin{center}
\epsfig{file=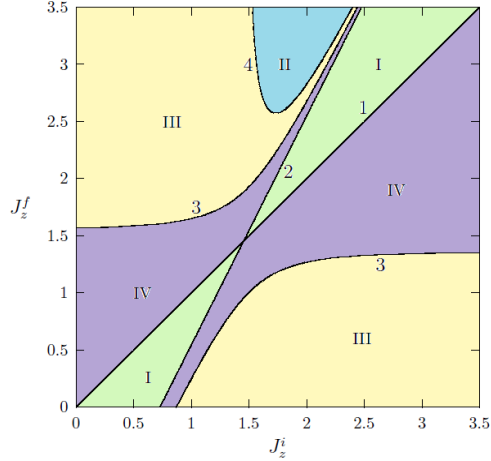,width=6.8cm}
\end{center}
\begin{center}
\caption{
(Color online)
Operation modes for a quantum Otto cycle in the plane $J_z^i$-$J_z^f$.
The regions corresponding to each operation mode are marked as
I,~heat engine - green; II,~refrigerator - blue; III,~heater - yellow; and
IV,~accelerator - violet.
Lines 1 and 2 are the boundaries $W=0$, while the curves 3 and 4 result from
conditions $Q_h=0$ and $Q_c=0$, respectively.
Straight lines 1 and 2 are the boundaries $W=0$, and curves 3 and 4 follow from the
conditions $Q_h=0$ and $Q_c=0$, respectively.
Parameters $R_1=3$ and $R_2=0.05$.
Temperatures of thermal reservoirs are $T_c=1$ and $T_h=2$
}
\label{fig:z22a}
\end{center}
\end{figure}

A similar scheme of regions for the operating modes is shown in Fig.~\ref{fig:z22a}.
%
%
It corresponds to the following parameter values: $R_1=3$, $R_2=0.05$, $T_c=1$, and
$T_h=2$.
According to Eqs.~(\ref{eq:Jzf_w1}) and (\ref{eq:Jzf_w}), the boundaries 1 and 2
intersect at the point $(1.45295,1.45295)$.
Above this point, the total work output has minimumal value of $W=-0.044432$ at the
point $(2.79285,3.35601)$.
The heat $Q_h$ at this point equals 0.168719.
Therefore the efficiency of heat engine is 26.3$\%$.
Moreover, below of the intersection point, the work $W(J_z^i,J_z^f)$ has the second
local minimum.
It is located at $(-0.104884,-0.762864)$ and equals $W=-0.119575$.
Here $Q_h=0.63495$ and hence $\eta_{mp}=18.8\%$.
Both these efficiencies are less than $\eta_N=29.3\%$.

\begin{figure}[t]
\begin{center}
\epsfig{file=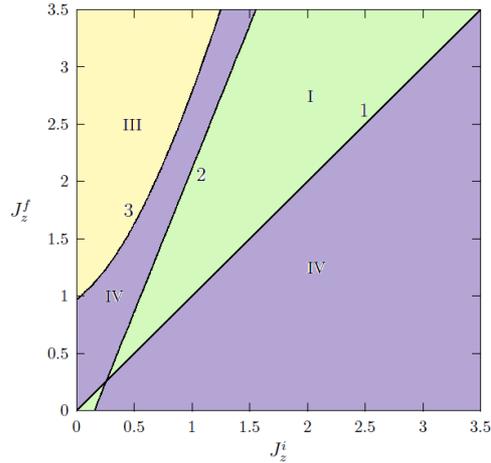,width=6.8cm}
\end{center}
\begin{center}
\caption{
(Color online)
The same as in Figs.~\ref{fig:z21} and \ref{fig:z22a}, but for
$R_1=1.3$, $R_2=0.8$, $T_c=1$ and $T_h=2.5$
}
\label{fig:z11}
\end{center}
\end{figure}

Next, Fig.~\ref{fig:z11} shows the operating mode areas for the following parameters:
$R_1=1.3$, $R_2=0.8$, $T_c=1$, and $T_h=2.5$.
%
%
The picture here is similar to the previous two cases.

Concluding this subsection, we can state the following.
Only four different operating modes were observed for the thermal machine under
study.
They are listed in Table~\ref{tab:4modes1}.
%
\begin{table}[t]
\caption{
Operating modes of the Otto machine depending on signs
$Q_c$, $W$ and $Q_h$
\upshape\upshape}
\label{tab:4modes1}
\begin{tabular}{lllll}
\hline\noalign{\smallskip}
mode & $Q_c$ & $W$ & $Q_h$ & scheme \\
\noalign{\smallskip}\hline\noalign{\smallskip}
engine & $-$ & $-$ & $+$ & $\{\leftarrow\uparrow\leftarrow\}$ \\
refrigerator & $+$ & $+$ & $-$ & $\{\rightarrow\downarrow\rightarrow\}$ \\
heater   & $-$ & $+$ & $-$ & $\{\leftarrow\downarrow\rightarrow\}$ \\
accelerator & $-$ & $+$ & $+$ & $\{\leftarrow\downarrow\leftarrow\}$ \\
\noalign{\smallskip}\hline
\end{tabular}
\end{table}
%
%
Although there are eight ($2^3=8$) different combinations of signs for $Q_c$, $W$ and
$Q_h$, the regimes
$\{\rightarrow\downarrow\leftarrow\}$ and $\{\leftarrow\uparrow\rightarrow\}$
are prohibited by the first law of thermodynamics (\ref{eq:WQ1}).
Moreover, as noted in Ref.~\cite{ZLCL07}, the variants ($Q_c>0, W>0, Q_h<0$) and
($Q_c>0, W>0, Q_h>0$),  or in our notation
$\{\rightarrow\uparrow\rightarrow\}$ and $\{\rightarrow\uparrow\leftarrow\}$,
contradict the second law of thermodynamics ($\oint\delta Q/T\ge0$ or $dS\ge0$).
As seen from Figs.~\ref{fig:z21}--\ref{fig:z11}, the operating mode regions
alternate in the following order: engine-accelerator-heater-refrigerator.

\section{
Concluding remarks
}
\label{sect:concl}
In the present paper, we have examined a two-qubit Heisenberg XYZ model with DM and
KSEA interactions under a non-uniform external magnetic field as the working
 substance
of a quantum Otto thermal machine.
Equations~(\ref{eq:Ei}) and (\ref{eq:R1R2}) show, firstly, that the KSEA interaction
affects the operation of the machine only through the collective parameter $R_1$, and
DM interaction only through $R_2$, and secondly, the roles of DM and KSEA interactions
change places when the longitudinal exchange constant $J_z$ changes the
antiferromagnetic behavior to ferromagnetic.

Combining analytical and numerical analysis, we have found regions in the parameter
space for possible operating modes of the thermal machine.
Only such four modes as a heat engine, refrigerator (heat pump), heater or dissipator
(when work is converted into the heat of both baths at once) and a thermal accelerator
or cold-bath heater
 (fast defrost regime)
are acceptable.

The engine and refrigerator mode regions can directly border each other
(Fig.~\ref{fig:zr1_2d})
or they are separated by areas with accelerator and heater regimes
(Figs.~\ref{fig:z21} and \ref{fig:z22a}).

We have found and investigated the efficiency of the heat engine at maximum
output power.
Remarkably, cases have been discovered where there are two local extrema of the total
work; their appearance is due to splitting the engine mode region into two subregions.
Optimal efficiency has been observed not only less than the Novikov efficiency, but
also greater than it for certain choices of model parameters.
However, the Carnot efficiency was never exceeded.

\vspace{-10mm}
\section*{}
{\bf Acknowledgment}\ 
%
Two of us, E.~K. and M.~Yu.,
were supported
 by
the program
CITIS \#AAAA-A19-119071190017-7.

%



\end{document}